\documentclass{emulateapj}
\usepackage{graphicx,amsfonts,natbib}
\shorttitle{GMCs in M31}
\shortauthors{Rosolowsky}
\begin{document}
\title{Giant Molecular Clouds in M31 -- I: Molecular Cloud Properties}
\author{E. Rosolowsky\altaffilmark{1}}
\affil{Center for Astrophysics, 60 Garden St. MS-66, Cambridge, MA 02138}
\email{erosolow@cfa.harvard.edu}
\altaffiltext{1}{National Science Foundation Astronomy and Astrophysics
Postdoctoral Fellow}

\begin{abstract}
We present Berkeley Illinois Maryland Association (BIMA) millimeter
interferometer observations of giant molecular clouds (GMCs) along a
spiral arm in M31.  The observations consist of a survey using the
compact configuration of the interferometer and follow-up,
higher-resolution observations on a subset of the detections in the
survey.  The data are processed using an analysis algorithm designed
to extract GMCs and correct their derived properties for observational
biases thereby facilitating comparison with Milky Way data.  The
algorithm identifies 67 GMCs of which 19 have sufficient
signal-to-noise to accurately measure their properties.  The GMCs in
this portion of M31 are indistinguishable from those found in the
Milky Way, having a similar size-line width relationship and
distribution of virial parameters, confirming the results of previous,
smaller studies.  The velocity gradients and angular momenta of the
GMCs are comparable to the values measured in M33 and the Milky Way;
and, in all cases, are below expected values based on the local
galactic shear.  The studied region of M31 has a similar interstellar
radiation field, metallicity, Toomre $Q$ parameter, and midplane
volume density as the inner Milky Way, so the similarity of GMC
populations between the two systems is not surprising.
\end{abstract}

\section{Introduction}
As the instrumentation for millimeter-wave telescopes improves, it
becomes progressively more straightforward to study individual
molecular clouds in other galaxies.  Recent studies of Local Group
galaxies have surveyed large numbers of molecular clouds in the Large
Magellanic Cloud \citep{nanten}, the Small Magellanic Cloud
\citep{nanten-smc}, M33 \citep{eprb03}, and a bevy of Local Group
dwarf galaxies \citep[e.g.~][]{wilson_6822,taylor_1569}.  These recent
studies explore the nature of star formation on galactic scales by
studying the properties of giant molecular clouds (GMCs,
$M>10^5~M_{\odot}$) throughout their host galaxies.  Such GMCs contain
the majority of the molecular mass in the Milky Way's ISM and are
responsible for most of the star formation in the Galaxy \citep{psp3}.

The Andromeda Galaxy (M31) is the second largest disk galaxy in the
Local Group, after the Milky Way, and it subtends over 2 deg$^{2}$ on
the sky.  Its proximity \citep[770 kpc,][]{m31-dist} makes it an
excellent target for studying extragalactic molecular clouds.
Numerous surveys of CO emission have been conducted over a portion of
M31 and a comprehensive list of the 24 CO studies published up to 1999
is given in \citet{fcrao-m31}.
This extensive list of surveys can be supplemented with a few major
studies that have occurred since then.  \citet{sheth,sheth06} used the
BIMA millimeter interferometer to study a $3'$ field in the outer
region of the galaxy ($R_{gal}=12$~kpc) and find 6 molecular complexes
similar to those found in the Milky Way. An extensive survey covering
the entirety of the star-forming disk of M31 has been completed using
the IRAM 30-m by \citet[][see also references therein]{iram-m31-aa}.
Finally, \citet{muller-thesis} used the Plateau de Burre
interferometer to examine the properties of molecular clouds in 9
fields.  Using the GAUSSCLUMPS \citep{gaussclumps,gaussclumps2}
algorithm, they decompose the emission into 30 individual molecular
clouds.


Previous high-resolution observations of CO in M31 indicate that a
large fraction of the molecular gas is found in GMCs.  Identifying
individual GMCs requires a telescope beam with a projected size
$\lesssim 50$ pc, the typical size of a GMC in the Milky Way
\citep{psp3}, which requires an angular resolution of $14''$ at the
distance of M31.  There have been seven observational campaigns that
observed CO $(1\to 0)$ emission from M31 at sufficient resolution to
distinguish molecular clouds: \citet{ich85, vbb87, lada_m31,
wilson_m31, loinard-ovro, sheth, muller-thesis}.  With the exception
of \citet{loinard-ovro}, all of these studies have found GMCs with
properties similar to those found in the inner Milky Way and
\citet{iram-bulge} have argued that the differences observed by
\citet{loinard-ovro} can be attributed to observational errors.
Indeed, \citet{vbb87} presented the first direct observations of GMCs
in any external galaxy using interferometric observations.  Subsequent
studies with interferometers and single-dish telescopes confirmed that
most CO emission in M31 comes from GMCs and that the GMCs properties
were similar to those found in the Milky Way \citep{lada_m31,
wilson_m31, sheth, muller-thesis}.

Although the molecular gas in M31 has been extensively studied, there
remains a gap connecting the large-scale, single-dish observations and
the small-scale, interferometer observations.  To address this gap, we
completed CO($J=1\to 0$) observations of a large (20~kpc${^2}$ region)
along a spiral arm of M31 with high resolution ($\sim 50$ pc).  We
then followed up on these observations using a more extended
configuration of the interferometer yielding data with a resolution of
$\sim 25$ pc.  This paper presents the observational data of the both
the survey and the follow-up observations (\S\ref{obs}).  Using only
the follow-up data, we present the first results, namely a
confirmation of previous studies that find GMCs in M31 are similar to
those in the Milky Way (\S\S \ref{analysis},\ref{larson-sec}).  Notably,
this paper utilizes the techniques described in \citep{props} to
correct the observational biases that plague extragalactic CO
observations, thereby placing derived cloud properties on a common
scale that can be rigorously compared with GMC data from other
galaxies.  The follow-up observations are also used to examine the
velocity gradients and angular momentum of the GMCs, which are then
compared to the remainder of gas in the galaxy for insight into the
GMC formation problem (\S\ref{spang}).  We conclude the paper by
examining the larger galactic environment of M31 to explore
connections between the GMCs and the larger ISM (\S\ref{environment}).
Subsequent work will explore the star formation properties of these
GMCs and the formation of such clouds along the spiral arm using the
data from the spiral arm survey.

\section{Observations}
\label{obs}
\begin{figure}
\plotone{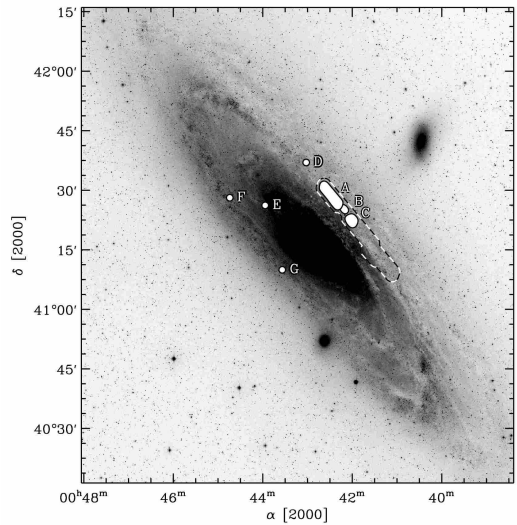}
\caption{\label{fields-m31} The Digital Sky Survey image of M31 with
the locations of the interferometer observations indicated.  The
extent of the D-array spiral arm survey is indicated with the black
and white dashed contour.  The targets of the C-array follow-up
observations are indicated with white regions.  Letters next to the
C-array fields indicate the corresponding fields in Table
\ref{m31-obsprops}.}
\end{figure}

We observed $^{12}$CO($J=1\to 0$) emission from M31 during the spring
and fall observing seasons of 2002 with the D and C configurations of
the BIMA millimeter interferometer \citep{bima}.  The observations
consisted of an 81-field mosaic using the most compact (D)
configuration with follow-up observations on seven sub-regions,
covering 30 fields at higher resolution (C array).  The D-array survey
spans a projected length of 6.7 kpc along a spiral arm in the
galaxy. Three of the seven follow-up, C-array fields targeted regions
with known CO emission from the D-array survey, and the remaining four
fields targeted regions with strong CO emission in the single-dish
observations of \citet{dame-m31} over a range of galactocentric
distances.  The locations of the fields are indicated in Figure
\ref{fields-m31}.

The D-array observations were completed in September and October 2002
over the course of four nights.  Each night roughly 20 pointings of
the mosaic were observed.  During the observations, the fields were
observed for 46 seconds each, making two passes through the mosaic
before returning to the phase calibrator (0102+504, 2.6 Jy) every 30
minutes.  This cycle was continued through the night, accumulating
$\sim 6$ hours of integration time on M31 per night (18 minutes per
field).  The correlator was configured to span 200 MHz at 500 kHz (1.3
km s$^{-1}$) resolution, easily encompassing the whole range of
velocities expected from the M31's rotation curve for this region
\citep{m31-rotcurve}.  At the latitude of the Hat Creek Radio
Observatory, M31 transits near zenith and cannot be observed for 40
minutes a night.  This time was used to perform flux calibrations
using Saturn and Uranus.  Over the course of the observations, the
derived flux of the phase calibrator was stable to 10\%.  The relative
flux calibration of the data may be better than this level owing to
the intrinsic variability of the phase calibrator.

The data were reduced and inverted using the MIRIAD software package
\citep{miriad}, following the calibration procedures of the BIMA
Survey of Nearby Galaxies \citep{song2}.  The four nights of $uv$ data
were combined using a linear mosaicing technique.  The inversion used
natural weighting, and gridding in the $uv$ plane was chosen to
produce maps with $3''$ pixels and $2.03$~km~s$^{-1}$ channel width.
The dirty maps were deconvolved with a Steer-Dewey-Ito deconvolution
algorithm optimized for mosaics (MOSSDI2 in the MIRIAD software
package).  Each plane of the final image was cleaned to the
1.5$\sigma_{rms}$ significance level.  The final resolution of the
survey is $14''$, with negligible variation over the map.   The mosaic
is corrected for the gain of the primary beam.  

Three regions within the D-array survey containing emission were
observed with the more extended C-array configuration.  In total, 26
fields were observed on five nights from October to December 2002.
The C-array observations used the same observing strategy as the
low-resolution survey, observing each field for 46 seconds and making
two or three passes (depending on the size of the mosaic) through the
mosaic before observing the phase calibrator every 24 minutes.  The
correlator configuration and choice of calibrators were the same as
for the survey.  The follow-up observations used a subset of the
D-array pointing centers to facilitate merging the two data sets.
These data were combined with the D-array observations in the $uv$
plane and inverted using uniform weighting.  The data were cleaned
using the same Steer-Dewey-Ito deconvolution algorithm as for the
D-array map.  The final resolution of the combined data is $\sim 9''$
with a 2.03~km~s$^{-1}$ channel width.  The images
are corrected for the gain of the primary beam.  

Prior to the D-array survey, four fields of M31 were observed using
the C array (Spring 2002).  These observations targeted single fields
known to contain bright CO emission from \citet{dame-m31} over a range
of galactocentric distance.  The fields were chosen to span a variety
of regions in the galaxy (different spiral arms and galactocentric
radii) in an attempt to reveal any biases seen by focusing on one
region of the galaxy.  The observations used 0136+478 as a phase
calibrator and Mars to establish the flux scale of the phase
calibrator.  Since the observations consisted exclusively of C-array
data, they were inverted using natural weighting.  The final
synthesized beam size for these observations is $7.5''$.

\begin{figure*}
\plotone{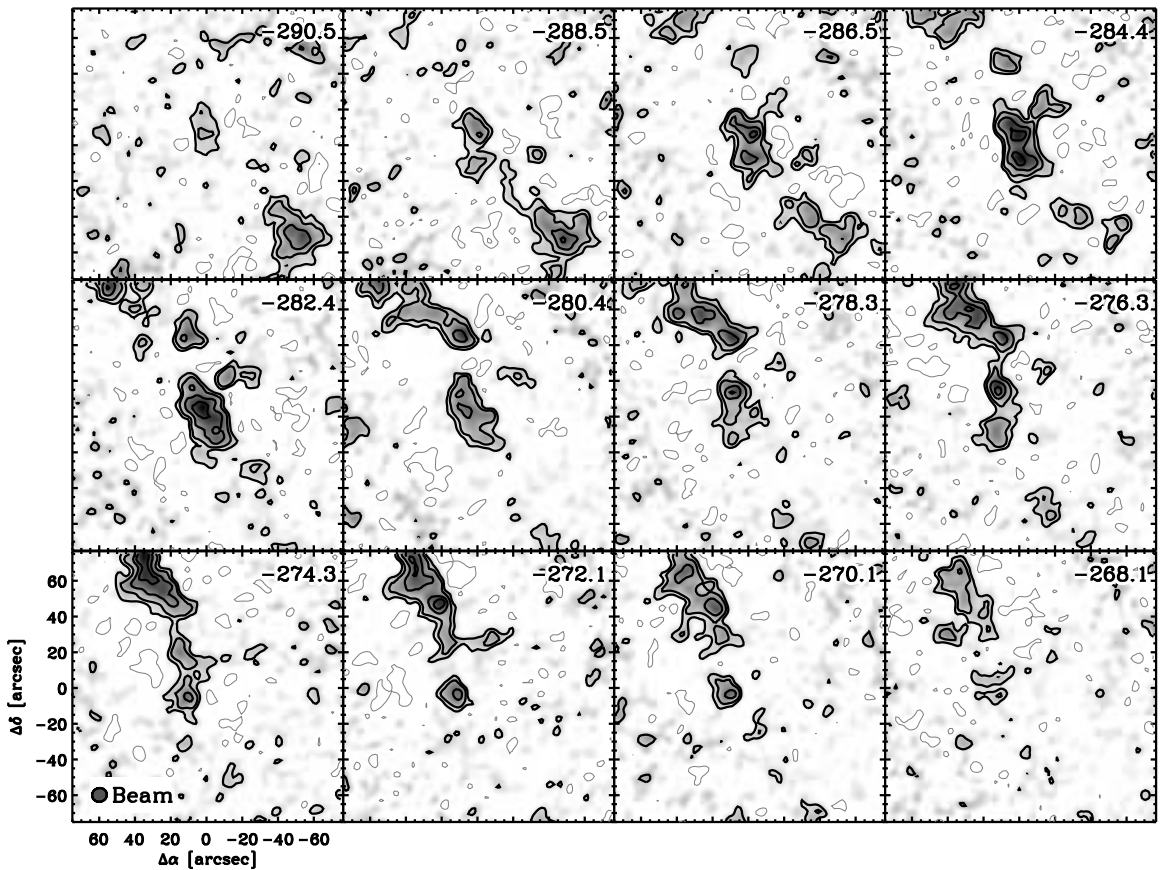}
\caption{\label{data-example} Channel maps of an excerpt from the
center of Field A.  Only 12 channel maps containing significant CO
emission are shown.  The gray scale of the channel map runs linearly
from 0 to 6 K~km~s$^{-1}$.  Contours of local significance are shown
at $-4,-2$ (gray),2, 4, 8, and 12 times the local value of
$\sigma_{rms}$.  At the center of the field, $\sigma_{rms}=0.4$
K~km~s$^{-1}$, though this value changes across the map owing to
variable gain from the interferometer. Each channel map is annotated
in the upper, right-hand corner with $V_{\mathrm{LSR}}$ for the
channel in km~s$^{-1}$. For
reference, the center of the field is at
$\alpha_{2000}=00^{\mathrm{h}}~42^{\mathrm{m}}~29\fs 3$,
$\delta_{2000}=+41^{\circ}~28'~53\farcs 3$ and the extent of the displayed
region is indicated in Figure \ref{m31-cfield1}.}
\end{figure*}

The observations produced one survey data cube and 7 high-resolution
data cubes for analysis.  The properties of these data sets are
summarized in Table \ref{m31-obsprops}.  Since space constraints
prohibit a full presentation of all the observational data, an excerpt
from the center of Field A is shown in Figure \ref{data-example} to
illustrate typical quality of the data.  The remainder of the paper
emphasizes the high-resolution data (Fields A-G in Table
\ref{m31-obsprops}).  We defer presentation and analysis of the Survey
data to a subsequent paper.

\begin{deluxetable*}{ccccccc}
\tablecolumns{6}
\tablewidth{0pt}
\tabletypesize{\footnotesize}
\tablecaption{\label{m31-obsprops} Summary of BIMA observations of M31.}
\tablehead{
\colhead{Field} & \colhead{Conf.} & \colhead{Center} &
\colhead{Resolution} & \colhead{Size} & \colhead{Noise} & \colhead{$f_{rec}$\tablenotemark{1}}\\
&& \colhead{($\alpha_{2000},\delta_{2000}$)} &
\colhead{($''\times \mbox{km/s}$)} & 
\colhead{($''\times ''\times \mbox{km/s}$)} & 
\colhead{(K km/s)} & }
\startdata
Survey & D & 00 41 48.1, +41 19 02 & 
$14\times 2.03$ & $1885\times 350\times 215$ & 0.27 & 0.35 \\
 A & C,D & 00 42 26.5,  +41 28 45 & 
$8.2\times 2.03$ & $255\times 205 \times 102$ & 0.64 & 0.35\\
 B & C,D & 00 42 12.3, +41 25 33
& $10.3\times 2.03$ & $570 \times 320 \times 102$ & 0.54 & 0.12\\
 C & C,D & 00 42 02.4,  +41 22 13
& $10.2\times 2.03$ & $305 \times 250 \times 102$ & 0.76 & 0.17 \\
 D & C & 00 43 01.8,  +41 37 00
& $7.0\times 2.03$ & $100\times 100 \times 100$ & 0.48 & 0.42 \\
 E & C &  00 43 56.8,  +41 26 12
& $8.1\times 2.03$ & $100\times 100 \times 100$ & 0.94 & 0.23 \\
 F & C & 00 44 44.3,  +41 28 07
& $7.2\times 2.03$ & $100\times 100 \times 100$ & 0.92 & 0.15 \\
 G & C & 00 43 33.9,  +41 09 58
& $7.1\times 3.04$ & $100\times 100 \times 100$ & 1.38 & 0.19\\
\enddata
\tablenotetext{1}{The fraction of flux recovered relative to the IRAM
30-m map of the galaxy \citep{iram-m31-aa}.}
\end{deluxetable*}

\section{Analysis of the Molecular Emission}
\label{analysis}

\subsection{Signal Identification}
\label{m31-masking}
For data cubes A--G in Table \ref{m31-obsprops}, we identified the
emission using a dynamic thresholding technique.  For every pixel in
the cube, we first estimated the rms value for the noise using the
method of \citet{eprb03}; this method accounts for gain variations in
both position and frequency.  We first estimate the noise at every
position by measuring the standard deviation of the data values using
iterative rejection of high signal pixels ($> 3\sigma_{rms}$).  We
generate a noise map by smoothing the resulting pixel estimates with a
boxcar kernel of the same size as the synthesized beam.  Then, we
estimate the relative changes of the noise across the bandpass by
estimating $\sigma_{rms}$ in each channel map and normalizing to the
mean value of $\sigma_{rms}(v)$ across the bandpass.  We generate a
cube of noise estimates by scaling the error map by the fractional
change from the mean appropriate for that velocity.  Performing a
pixel-wise division of the data cube by the cube of noise estimates
yields a data set in significance units.

We identify CO emission in the data cubes by searching for contiguous
regions of high significance in position-position-velocity space using
a two-tiered thresholding scheme \citep[e.g.~][]{syscw}.  The core
pixels in the mask are identified as pairs of pixels in adjacent
velocity channels with $I(x,y,v) > 4\sigma_{rms} (x,y,v)$.  The mask
is then expanded in position and velocity space to include all pixels
with $I > 2 \sigma_{rms}$ that are connected by high significance
pixels to the $4\sigma_{rms}$ core of the mask.  This masking process
has been used in previous extragalactic, interferometric studies and
has been shown to successfully identify emission with minimal
inclusion of noise \citep{eprb03}.  Eliminating noise from the
emission under consideration is key to the methods used to determine
cloud properties (see \S\ref{cloudprops}).

\subsection{Inclination Effects}

The identification and analysis of GMCs in M31 may be complicated by
the relatively high inclination of the galaxy: $i=77^{\circ}$
\citep{wk88}.  The situation is
similar, in part, to observing GMCs in the outer regions of the Milky
Way ($i=90^{\circ}$!), and the spatial resolution ($\sim 25$ pc) is
comparable to what is found using small telescopes (e.g.~the CfA 1.2-m)
to observe distant ($>10$~kpc) GMCs in the Milky Way \citep{dht01}.
While study of GMCs has been conducted in the Milky Way with such
data, the inclination being $77^{\circ}$ and not $90^{\circ}$ helps
the situation significantly. The foreground and background spiral arms
in M31 are separated from the arm under consideration by $> 2'$ in
projection.  Thus, spatial information is useful for separating the
emission into GMCs.  The clouds in the interarm region are likely too
faint to be recovered by the deconvolution algorithm
(\S\ref{fluxrec}).  We conclude that the primary concern in
decomposition is the blending of GMCs within the same spiral arm.
\citet{iram-m31-aa} estimate that the width of the spiral arms in the
plane of the galaxy for this region is $\sim 500$~pc with a vertical
thickness of 150 pc.  There could be multiple, distinct clouds in the
spiral arm along the same line of sight, so we make a simple estimate
of how many GMCs are likely to be along a line of sight through the
spiral arm.  We model the arm as a cylinder with elliptical
cross-section having major and minor axes of 500 pc and 150 pc
respectively.  For a viewing angle 77$^{\circ}$ away from the minor
axis of the cylinder, the path length through the cylinder is $\ell =
400$~pc.  We assume that the GMCs are distributed uniformly in the
cylindrical volume with volume density $n$ and geometric cross-section
$\sigma$.  The mean number of clouds intersected by a line-of-sight
through the arm is then $n\sigma \ell$.  We estimate $n$ by counting
the number of clouds ($N$) in some trial volume along the arm $V = \pi
LWH/4$ where $L$,$W$ and $H$ represent the length, width (major axis)
and height (minor axis) of the volume of the arm respectively.  By
assuming a surface density through a typical GMC of $\Sigma_{GMC}$ and
comparing this to the observed surface density in the single-dish map
$\langle \Sigma_{\mathrm{H2}}\rangle$, we can estimate the area
filling fraction of GMCs in the arm when viewed from above the
galaxy:$f=\langle \Sigma_{\mathrm{H2}}\rangle/\Sigma_{GMC}$.  Since
$f$ is also given by $N\sigma/LW$, we can express
\begin{equation}
n\sigma\ell = \frac{4\ell \langle \Sigma_{\mathrm{H2}}\rangle}
{\pi H \Sigma_{GMC}}.
\end{equation}
Significant confusion will occur for $n\sigma \ell \gtrsim 1$.  For
this condition to be met with $\ell = 400$~pc and $H=150$~pc,
$\langle\Sigma_{\mathrm{H2}}\rangle/ \Sigma_{GMC} > 0.3$.  The data of
\citet{iram-m31-aa} show $\langle \Sigma_{\mathrm{H2}}\rangle\approx
5~M_{\odot}\mbox{ pc}^{-2}$ for this region requiring GMCs in M31 to
have $\Sigma_{GMC} < 20~M_{\odot}\mbox{ pc}^{-2}$ for significant
blending to occur. We regard this possibility as unlikely since GMCs
in other, less confused regions of M31 are found to have comparable
properties to Local Group clouds \citep{sheth,muller-thesis} where
$\Sigma_{GMC} \approx 100~M_{\odot}\mbox{ pc}^{-2}$
\citep{srby87,psp5}.  Unfortunately, one of the aims of the paper is
to assess how similar clouds in M31 are to those in other galaxies; so
we cannot reject the possibility that the clouds could be
significantly blended along the line of sight although this would
imply significantly larger sizes and lower column densities than is
typical for GMCs in other galaxies.

We conclude this discussion of inclination effects by noting that we
will assume that GMC properties are independent of viewing angle, as
is common in studies of GMCs in the Milky Way and beyond
\citep[e.g.][]{syscw,srby87,ws90,wilson_m31,sheth}.  As a result, the
reported properties of GMCs do not need any correction for the high
inclination of the galaxy.

\subsection{Identifying GMCs and Measuring Their Properties}
\label{cloudprops}

\begin{figure*}
\plotone{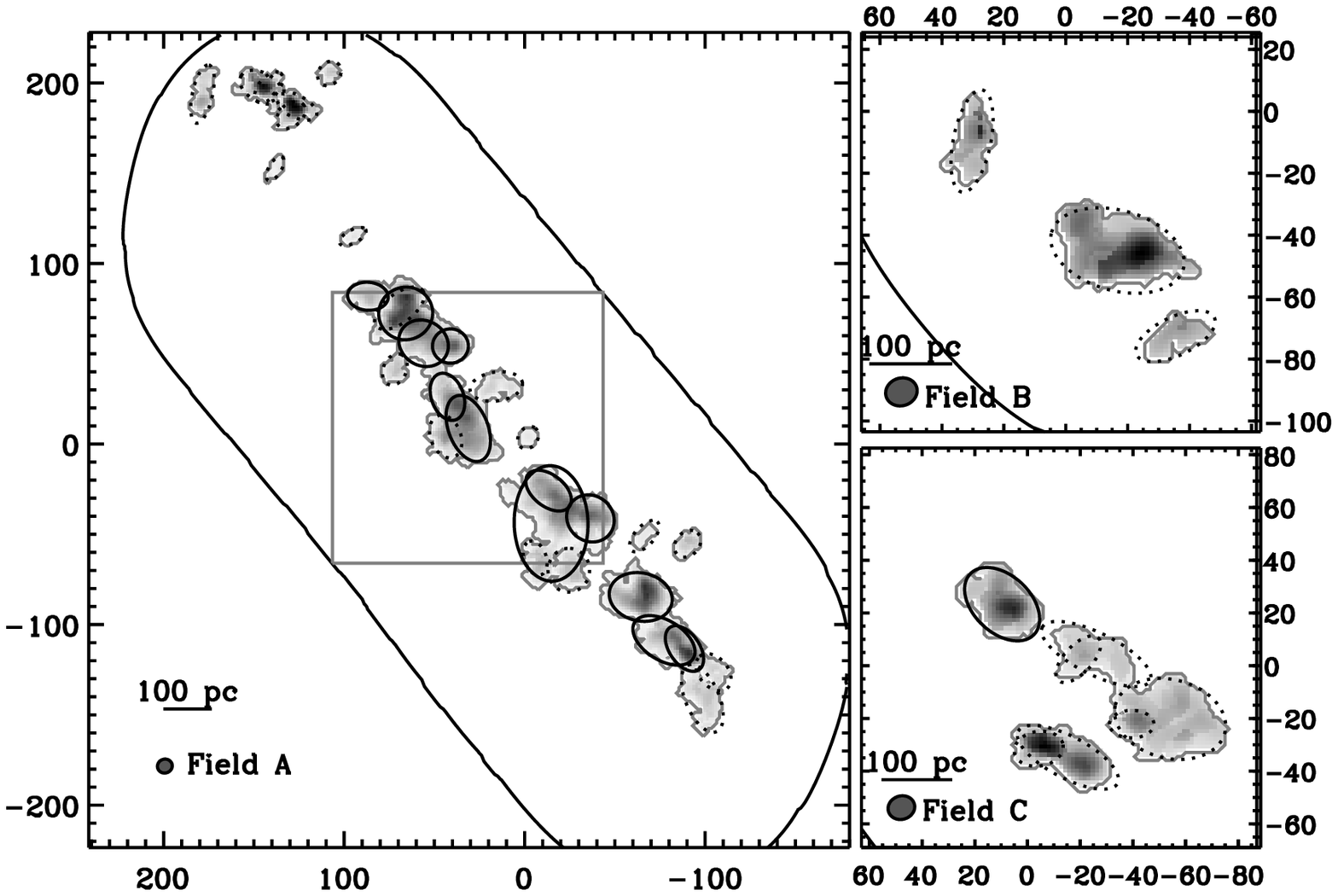}
\caption{\label{m31-cfield1} Masked integrated-intensity maps of high
resolution fields A--C.  The black ellipses indicated the
non-deconvolved major and minor axes of the molecular clouds
identified by the segmentation algorithm.  GMCs with sufficient signal
to noise to be included in the property analysis (\S\ref{larson-sec})
are shown with a solid black line and low signal-to-noise clouds are
shown with dotted ellipses.  The gray scale runs linearly from 0
K~km~s$^{-1}$ (indicated with a contour) to the maximum indicated in
the upper right-hand corner of each map.  Axes are aligned with
equatorial coordinates and are labeled in offsets measured in
arcseconds from the field centers listed in Tables \ref{m31-obsprops}.
For reference, a scale bar is plotted of projected distance in the
plane of the sky and the size of the synthesized beam is depicted
beside the field name.  If appropriate, the edge of the surveyed
region is drawn with a solid line.  The square, gray box in the map
for Field A indicates the region displayed in Figure
\ref{data-example}.}
\end{figure*}

\begin{figure*}
\plotone{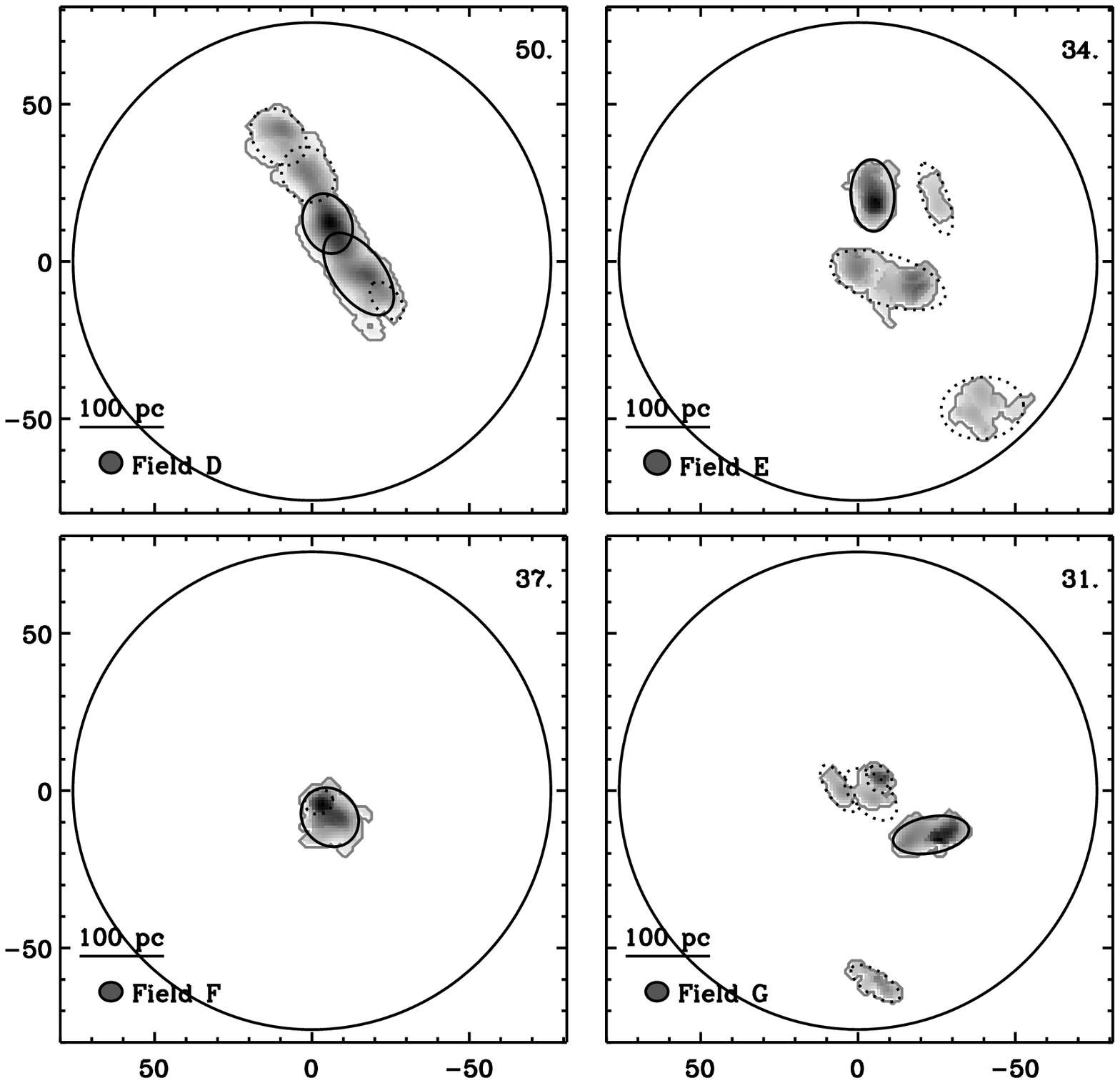}
\caption{\label{m31-cfield2} As Figure \ref{m31-cfield1} but for
Fields D--G.}
\end{figure*}

We identify emission in the data cubes from the high-resolution data
(Fields A-G in Table \ref{m31-obsprops}) using the
signal-identification method described in \S\ref{m31-masking}. Maps of
the high-resolution data fields appear in Figures \ref{m31-cfield1}
and \ref{m31-cfield2}.  This emission was partitioned into individual
molecular clouds using the segmentation\footnote{Segmentation is the
term used in image processing for dividing an image into subregions
that share common properties, in this case, being part of the same
GMC.} algorithm described in \citet[][RL06]{props} to separate blended
emission.  The segmentation algorithm is a modified watershed
algorithm, related to the original CLUMPFIND \citep{clumpfind}.
However, the adopted algorithm has been optimized to identify GMCs in
a wide variety of observational data: the algorithm is significantly
more robust to the presence of noise and does not over-segment GMCs as
the original CLUMPFIND is known to do \citep{sheth}.  The RL06
algorithm establishes its parameters from physical scales in the data
(e.g.~parsecs, km~s$^{-1}$) as opposed to observationally determined
scales (beam and channel widths) to minimize biases in comparing
catalogs with different observational properties.  For the
high-resolution data, the algorithm identifies 67 molecular clouds in
the seven cubes.


\begin{deluxetable*}{cccccccc}
\tablecolumns{8}
\tablewidth{0pt}
\tabletypesize{\footnotesize}
\tablecaption{\label{m31-cloudprops}Molecular Cloud Properties in M31}
\tablehead{
\colhead{Number} & \colhead{Position\tablenotemark{a}} & \colhead{$M_{LUM}$\tablenotemark{b}} & 
\colhead{$R_e$\tablenotemark{b,c}} & \colhead{$\sigma_v$\tablenotemark{b}} & \colhead{$T_{max}/\sigma_{rms}$} &
\colhead{$|\nabla v|$\tablenotemark{b}} & \colhead{$\phi_{\nabla}$\tablenotemark{b}} \\
\colhead{} & \colhead{$('','')$} & \colhead{($10^4~M_{\odot}$)} & \colhead{(pc)} & 
\colhead{(km s$^{-1}$)} & \colhead{} & \colhead{(km s$^{-1}$ pc$^{-1}$)} 
& \colhead{($^{\circ}$)}}
\startdata
\cutinhead{High Signal-to-Noise Clouds}
1 & $(+322,+1246)$ & 55 & 34 & 4.07 & 17.0 & 0.07 & $97$ \\
2 & $(+340,+1262)$ & 42 & 24 & 4.43 & 15.1 & 0.04 & $-151$ \\
3 & $(-469,+671)$ & 72 & 52 & 6.29 & 14.9 & 0.12 & $-66$ \\
4 & $(-238,+828)$ & 78 & 50 & 4.86 & 14.8 & 0.02 & $67$ \\
5 & $(-300,+764)$ & 63 & 47 & 3.38 & 14.6 & 0.03 & $132$ \\
6 & $(-256,+811)$ & 33 & 43 & 4.44 & 13.3 & 0.10 & $125$ \\
7 & $(+2384,+709)$ & 33 & 28 & 2.35 & 13.2 & 0.05 & $3$ \\
8 & $(-279,+782)$ & 25 & 32 & 5.45 & 12.6 & 0.11 & $39$ \\
9 & $(+1438,+623)$ & 31 & 19 & 2.86 & 12.3 & 0.03 & $106$ \\
10 & $(-379,+730)$ & 34 & 33 & 4.38 & 11.9 & 0.06 & $43$ \\
11 & $(-201,+837)$ & 15 & 24 & 2.38 & 11.5 & 0.06 & $58$ \\
12 & $(-381,+711)$ & 61 & 92 & 4.24 & 11.3 & 0.03 & $49$ \\
13 & $(-282,+810)$ & 26 & 27 & 4.07 & 11.3 & 0.08 & $150$ \\
14 & $(-276,+812)$ & 26 & 30 & 4.88 & 11.2 & 0.18 & $127$ \\
15 & $(-492,+647)$ & 26 & 47 & 2.95 & 10.7 & 0.03 & $126$ \\
16 & $(+948,-386)$ & 26 & 19 & 3.09 & 10.6 & 0.02 & $93$ \\
17 & $(-420,+714)$ & 42 & 42 & 4.41 & 10.6 & 0.06 & $112$ \\
18 & $(-512,+642)$ & 40 & 30 & 5.54 & 10.3 & 0.06 & $-50$ \\
19 & $(-820,+387)$ & 38 & 40 & 3.99 & 10.3 & 0.08 & $-145$ \\
\cutinhead{Low Signal-to-Noise Clouds}
20 & $(-254,+821)$ & 21 & 21 & 4.40 & 9.9 & 0.11 & $-166$ \\
21 & $(-864,+329)$ & 31 & 19 & 2.75 & 9.8 & 0.03 & $-29$ \\
22 & $(+368,+1290)$ & 27 & 25 & 5.39 & 9.7 & 0.08 & $64$ \\
23 & $(-234,+828)$ & 44 & 34 & 6.19 & 9.7 & 0.17 & $-88$ \\
24 & $(+351,+1278)$ & 24 & 24 & 5.89 & 9.3 & 0.11 & $41$ \\
25 & $(+1430,+596)$ & 45 & 35 & 4.67 & 9.2 & 0.11 & $26$ \\
26 & $(+978,-368)$ & 8 & \nodata & 4.64 & 8.8 & 0.16\tablenotemark{d} & $165$ \\
27 & $(-278,+760)$ & 22 & 28 & 4.13 & 8.7 & 0.09 & $32$ \\
28 & $(-586,+554)$ & 10 & \nodata & 2.62 & 8.4 & 0.01 & $112$ \\
29 & $(-129,+943)$ & 29 & \nodata & 6.53 & 8.2 & 0.21 & $111$ \\
30 & $(-199,+837)$ & 3 & \nodata & 1.56 & 8.1 & 0.05\tablenotemark{d} & $-28$ \\
31 & $(-399,+686)$ & 17 & 37 & 4.70 & 8.0 & 0.13 & $53$ \\
32 & $(-530,+615)$ & 23 & 42 & 4.51 & 7.8 & 0.07 & $172$ \\
33 & $(-96,+953)$ & 38 & 27 & 7.03 & 7.7 & 0.14 & $-172$ \\
34 & $(-38,+949)$ & 15 & \nodata & 2.38 & 7.6 & 0.04 & $-35$ \\
35 & $(-327,+787)$ & 12 & 29 & 3.15 & 7.5 & 0.05 & $-28$ \\
36 & $(-125,+938)$ & 14 & \nodata & 3.93 & 7.4 & 0.09 & $103$ \\
37 & $(-285,+780)$ & 17 & 19 & 5.26 & 7.2 & 0.18 & $102$ \\
38 & $(-232,+831)$ & 12 & 31 & 2.67 & 7.1 & 0.04 & $-154$ \\
39 & $(-669,+519)$ & 37 & 55 & 2.64 & 6.9 & 0.02 & $-84$ \\
40 & $(+2390,+714)$ & 3 & \nodata & 1.50 & 6.7 & 0.07\tablenotemark{d} & $-133$ \\
41 & $(+307,+1238)$ & 2 & \nodata & 2.42 & 6.7 & 0.06\tablenotemark{d} & $-134$ \\
42 & $(-933,+343)$ & 33 & 60 & 2.97 & 6.6 & 0.04 & $39$ \\
43 & $(-907,+342)$ & 6 & \nodata & 3.62 & 6.6 & 0.05\tablenotemark{d} & $-57$ \\
44 & $(-844,+332)$ & 12 & \nodata & 3.65 & 6.5 & 0.06\tablenotemark{d} & $134$ \\
45 & $(-541,+629)$ & 9 & 25 & 3.42 & 6.4 & 0.04 & $179$ \\
46 & $(+1000,-371)$ & 5 & \nodata & 1.58 & 6.3 & 0.02\tablenotemark{d} & $-79$ \\
47 & $(+1376,+556)$ & 16 & 35 & 2.08 & 6.3 & 0.04 & $100$ \\
48 & $(-227,+797)$ & 8 & \nodata & 3.51 & 6.1 & 0.12 & $18$ \\
49 & $(-515,+700)$ & 10 & \nodata & 2.97 & 6.1 & 0.11 & $-29$ \\
50 & $(-274,+792)$ & 3 & \nodata & 1.74 & 6.1 & 0.06\tablenotemark{d} & $-27$ \\
51 & $(-235,+837)$ & 5 & 12 & 2.16 & 6.1 & 0.06\tablenotemark{d} & $168$ \\
52 & $(-703,+491)$ & 6 & \nodata & 2.24 & 6.1 & 0.05 & $68$ \\
53 & $(-186,+870)$ & 2 & \nodata & 1.70 & 6.0 & 0.06\tablenotemark{d} & $-124$ \\
54 & $(-366,+692)$ & 10 & \nodata & 6.27 & 5.9 & 0.15 & $77$ \\
55 & $(+980,-433)$ & 8 & \nodata & 5.00 & 5.9 & 0.14\tablenotemark{d} & $-154$ \\
56 & $(+982,-373)$ & 6 & 16 & 2.14 & 5.8 & 0.05\tablenotemark{d} & $-68$ \\
57 & $(-871,+366)$ & 3 & \nodata & 1.49 & 5.5 & 0.00\tablenotemark{d} & $73$ \\
58 & $(-225,+801)$ & 3 & \nodata & 2.60 & 5.2 & 0.08\tablenotemark{d} & $79$ \\
59 & $(-227,+794)$ & 4 & \nodata & 2.38 & 5.1 & 0.04\tablenotemark{d} & $46$ \\
60 & $(-110,+908)$ & 2 & \nodata & 1.79 & 5.1 & 0.04\tablenotemark{d} & $-8$ \\
61 & $(-359,+759)$ & 1 & \nodata & 0.99 & 5.1 & 0.01\tablenotemark{d} & $147$ \\
62 & $(-163,+961)$ & 4 & \nodata & 2.42 & 4.9 & 0.11\tablenotemark{d} & $13$ \\
63 & $(-184,+776)$ & 1 & \nodata & 0.98 & 4.8 & 0.04\tablenotemark{d} & $-59$ \\
64 & $(-473,+704)$ & 3 & \nodata & 1.28 & 4.6 & 0.02\tablenotemark{d} & $-100$ \\
65 & $(-883,+368)$ & 14 & 22 & 3.11 & 4.6 & 0.04 & $115$ \\
66 & $(+1403,+622)$ & 3 & \nodata & 0.65 & 4.5 & 0.01\tablenotemark{d} & $-57$ \\
67 & $(-103,+953)$ & 4 & \nodata & 2.58 & 4.2 & 0.12\tablenotemark{d} & $-6$ \\

\enddata
\tablenotetext{a}{Position given in arcseconds
relative to the center of M31 at
$\alpha_{2000}=00^{\mathrm{h}}~42^{\mathrm{m}}~44\fs 3$ and
$\delta_{2000}=+41^{\circ}~16'~09''$}
\tablenotetext{b}{Typical errors in properties --- Luminous Mass
($M_{LUM}$): 20\%; Radius ($R_e$):15\%; Line Width ($\sigma_v$):15\%,
Velocity Gradient Magnitude($|\nabla v|$):50\%,  Gradient Position
Angle ($\phi_{\nabla}=30^{\circ}$).}
\tablenotetext{c}{No values of the deconvolved radius ($R_e$) are reported if
the cloud cannot be resolved with the present data.}
\tablenotetext{d}{The uncertainty in the gradient of this cloud is 
larger than $0.15~\mbox{km s}^{-1}\mbox{ pc}^{-1}$ and it is not 
included in the gradient analysis.}
\end{deluxetable*}

\begin{figure}
\plotone{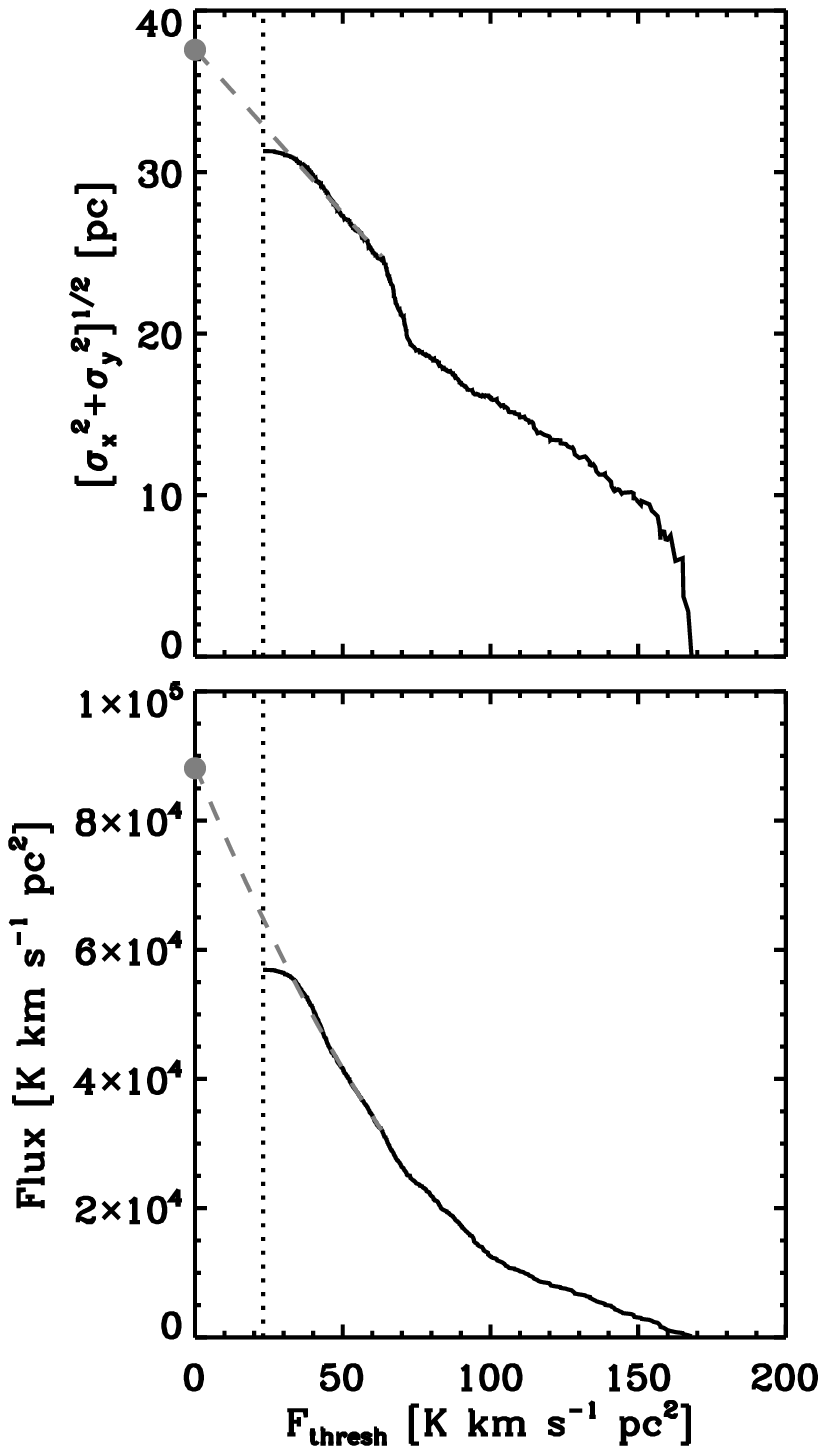}
\caption{\label{extrapfig} Extrapolation of cloud properties
({\it top}: non-deconvolved cloud radius, {\it bottom}: flux) to the
$F_{thresh}=0~\mbox{K km s}^{-1}\mbox{pc}^{2}$ flux threshold using
the extrapolation algorithm of RL06.  The extrapolation corrects for
signal lost below the 2$\sigma_{rms}=23~\mbox{K km
s}^{-1}\mbox{pc}^{2}$ clipping level (vertical dotted line).  The
extrapolated value is indicated with a gray dot and the extrapolation
is shown as a gray dashed line.  This demonstration of the
extrapolation is shown for Cloud 19 in Table
\ref{m31-cloudprops}, which is the lowest significance GMC
used in the analysis of cloud properties (\S\ref{larson-sec}).}
\end{figure}

For each of these clouds, we measure three macroscopic properties:
mass, radius ($R_e$), line width ($\sigma_v$).  These macroscopic
properties are defined using moments of the intensity distribution
(see RL06 for details).  The mass of the GMCs is determined by scaling
the zeroth moment (sum) of the intensity distribution by a constant
CO-to-H$_2$ conversion factor of
\[ X_{\mathrm{CO}}=2 \times 10^{20} \frac{\mbox{ cm}^{-2}}
{\mbox{ K km s}^{-1}}\]
\citep{sm96,dht01}.  Assuming a mean particle mass of
1.36$m_{\mathrm{H}}$ \citep{ism-abund} gives a mass from the
integrated flux over the cloud:
\[M_{LUM} = 4.4~M_{\odot}~\frac{W^*_{\mathrm{CO}}}
{\mbox{K km s}^{-1}\mbox{pc}^{2}}.\] The calculation uses a flux value
$W^*_{\mathrm{CO}}$ that is extrapolated from the brightness threshold
($2\sigma_{rms}$) to 0 $\mbox{K km s}^{-1}\mbox{pc}^{2}$ using
variation in $W_{\mathrm{CO}}$ as a function of threshold
\citep{syscw,props}.  RL06 demonstrate that this extrapolation 
practically eliminates bias in the data due to low sensitivity, though
incomplete flux recovery interferometer may still affect the resulting
mass estimates (\S\ref{fluxrec}).  An demonstration of the extrapolation
method used in this paper is shown in Figure \ref{extrapfig}.

In a similar fashion to the molecular mass, we use the moments of the
emission distribution to measure the sizes and line widths of the
resulting molecular clouds.  Following RL06, the size is defined as
the second moment of the emission distribution along the major and
minor axes of the cloud: $\sigma_{maj}$ and $\sigma_{min}$
respectively.  Like the total flux, the values of these moments are
extrapolated to the 0 K km s$^{-1}$ pc$^{2}$ intensity level to
minimize the effects of the brightness threshold giving
$\sigma_{maj}^*$ and $\sigma_{min}^*$ (see RL06 for details).  We
estimate the deconvolved size of the major and minor axes by
subtracting the width of the beam in quadrature.  The size of the
cloud is the geometric mean of the ``deconvolved'' major- and
minor-axis sizes:
\begin{equation}
\sigma_{r}=\left[\sqrt{(\sigma^*_{maj})^2-\sigma_{beam}^2}
\sqrt{(\sigma^*_{min})^2-\sigma_{beam}^2}\right]^{1/2}
\label{rdefn}
\end{equation} 
The radius of the cloud is estimated by scaling $\sigma_{r}$ up by a
factor of 1.91 \citep[][S87]{srby87}.  \citet{bertoldi-mckee} note
that this radius estimate is insensitive to inclination and projection
effects.

The rms velocity dispersion is calculated as the second moment of the
emission distribution in velocity space.  Again, the velocity
dispersion is extrapolated to to the the 0 K km s$^{-1}$ pc$^{2}$
intensity level to give the corrected velocity dispersion
$\sigma^*(v)$.  The properties of the molecular clouds in the
high-resolution data are summarized in Table \ref{m31-cloudprops}.
The clouds are divided into two groups: `high signal-to-noise clouds' with
$T_{peak}/\sigma_{rms}\ge 10$ where the macroscopic properties can be
accurately recovered (see \S\ref{fluxrec}) and  `low signal-to-noise clouds'
($T_{peak}/\sigma_{rms}< 10$) where the derived properties are suspect.

\subsection{Flux Recovery}
\label{fluxrec}
Since interferometers do not measure the total power of observed
emission, they must rely upon single dish observations for accurate
measurements of the CO flux.  In the absence of such single-dish data,
deconvolution algorithms can be used to extrapolate into the unsampled
short-spacing area of the $(u,v)$ plane \citep{helfer02}. The accuracy
of such extrapolations depends primarily on the signal-to-noise ratio
of the data and to a lesser degree on the observational strategy and
the structure of the object being observed.  Since this work uses flux
information to calculate cloud masses as well as in the moments that
determine the cloud properties, flux loss affects all stages of this
analysis.

We use the data from IRAM 30-m survey of M31 \citep{iram-m31-aa} to
quantify the amount of flux lost in each of the the observations.  We
report the fraction of flux in the BIMA maps relative to that found in
the IRAM map in Table \ref{m31-obsprops} as the column labeled
$f_{rec}$.  The fractions recovered range from 12\% to 42\%.  The
relatively low fraction of flux recovered is a product of two factors
that affect interferometric observations: spatial filtering of GMCs
\citep[studied in][]{sheth,sheth06} and the non-linear recovery of
signal in the low signal-to-noise regime \citep[explored
in][]{helfer02}.  The work of \citet{sheth,sheth06} found that,
because of their clumpy structure, spatial filtering did not hinder
the accurate measurement of GMC shapes, fluxes and properties.
However, their simulations were conducted in the high signal-to-noise
regime.  In the low signal-to-noise regime, \citet{helfer02} found
that even relatively compact simulated sources suffered from
significant flux loss, owing to the inability for the deconvolution
algorithms to isolate low-amplitude signal.  These two effects
(spatial filtering of GMC emission and low signal-to-noise) were
studied together in RL06 who simulated BIMA C,D, and C+D observations
of Milky Way GMCs (Orion, Rosette, W3/4/5) as if those GMCs were
located in M31.  Their work showed that accurate recovery of cloud
properties required signal-to-noise ratios of $T_{max}/\sigma_{rms}\ge
10$ for less than a 20\% loss in all the cloud properties.  For clouds
with lower significance, the flux loss is more severe.  However, the
dominant source of flux loss comes from clouds not detected in the
observations at all, i.e. those with $T_{max}/\sigma_{rms}\lesssim
1.5$, the minimum amplitude to which the deconvolution algorithm
operates.  For these clouds, the deconvolution algorithm does not
identify a significant local maximum for CLEANing and the cloud
contributes zero flux to the deconvolved map.  Such low mass clouds
likely comprise most of the missing flux not seen in the
interferometer data.  When clouds are detected, cloud properties
estimated from the observations should be relatively accurate,
particularly if $T_{max}/\sigma_{rms}\ge 10$. In general, we only
consider the properties of clouds in the study if this sensitivity
condition is met; exceptions to this criterion will be noted.  For the
high-resolution data, 19 GMCs meet the signal-to-noise criterion (the
`high signal-to-noise clouds' in Table \ref{m31-cloudprops}).  We
emphasize that for these M31 data interferometer observations act as a
filter that preferentially detects GMCs, but the properties of those
clouds are well-measured. 

\section{Molecular Cloud Properties in M31}
\label{larson-sec}

Upon measuring a set of GMC properties for clouds in another galaxy,
the immediate question is whether these clouds are similar to those
found in the Milky Way.  Ideally, the well-resolved clouds in the
Milky Way would be analyzed using the methods of RL06 so that the
results are directly comparable.  However, since measuring cloud
properties in the inner Galaxy requires establishing a single distance by
resolving the kinematic distance ambiguity, a detailed re-analysis has
not been completed to date.  The next best solution is to compare the
results to an existing catalog.

For comparison, we use the catalog of \citet[][S87]{srby87} to define
the population of GMCs in the inner Milky Way.  The work of S87 used
similar methods to calculate cloud properties though their work adopts
a fundamentally different procedure for segmenting GMCs.  In addition,
their cloud properties are not corrected for the effects of
sensitivity as are the properties derived in this study.
Consequently, there may be systematic differences between their
catalog and the present study.  In general, the fluxes, line widths
and radii of the S87 clouds will be smaller than would be derived
using the methods of RL06; however, using interferometer data to
observe the extragalactic clouds will reduce the fluxes, line widths
and radii by a similar amount (RL06). We have rescaled the catalog of
S87 to a different galactocentric solar radius: $R_{\odot}=10$~kpc to
$R_{\odot}=8.5$~kpc and the CO-to-H$_2$ conversion factor adopted in
this study.  We also recalculated cloud sizes using the definition of
RL06.  Finally, it is noteworthy that the method of RL06 reports, in
addition to unbiased values for the cloud properties, appropriate
uncertainties for those properties facilitating a statistically based
comparison of cloud properties in M31 to those in the Milky Way.

\subsection{Scalings Among GMC Properties}

\begin{figure*}
\plottwo{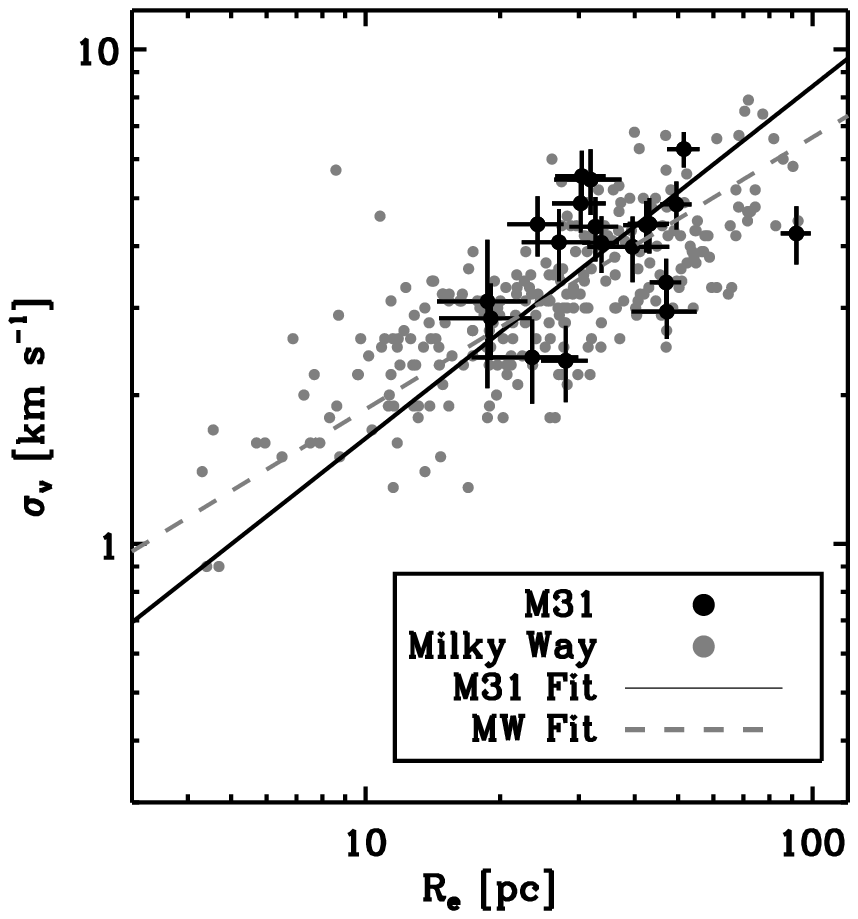}{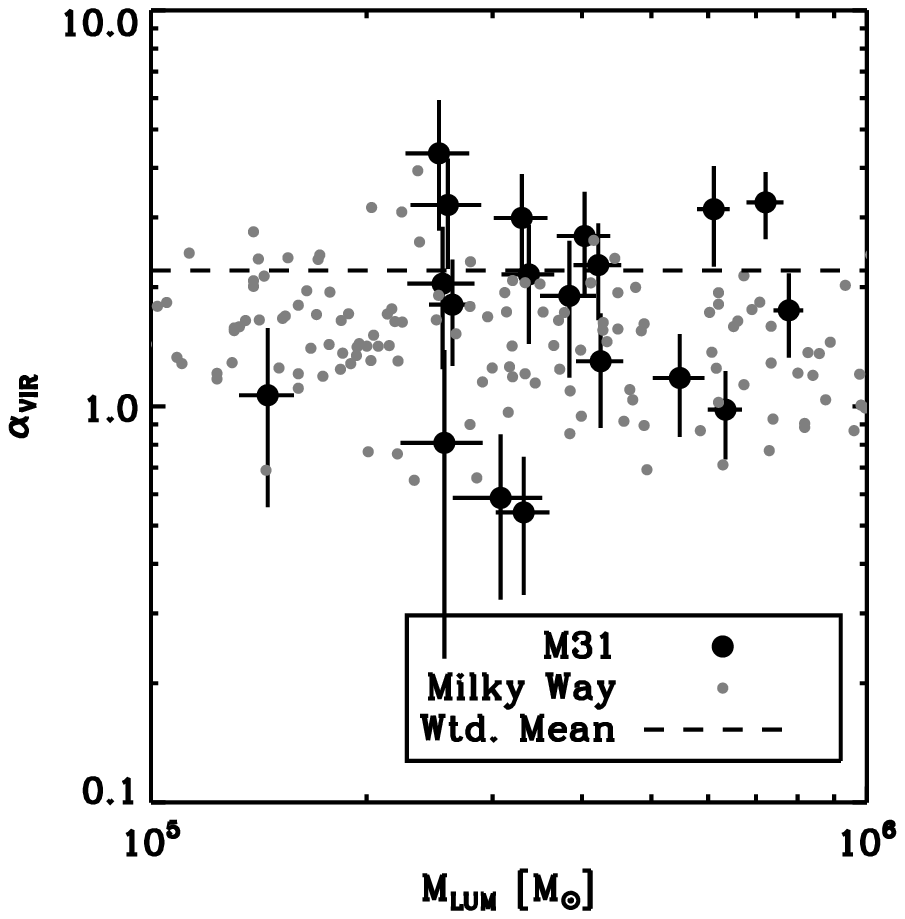}
\caption{\label{larson} ({\it left}) The size-line width relationship
for the 19 clouds in M31 with sufficient signal-to-noise for accurate
recovery of cloud properties.  There is good agreement between clouds
in M31 and those cataloged in the inner Milky Way by S87.  The solid,
black and dashed, gray lines are fits to the data for GMCs in M31 and
the inner Milky Way respectively; the fits are statistically
indistinguishable. ({\it right}) The virial parameter as a function of
luminous mass.  The virial parameter is roughly constant as a function
of luminous mass implying a linear scaling between the luminous and
virial masses of the clouds.}
\end{figure*}

\citet{larson} first demonstrated that there were power-law scalings
among the properties of molecular gas clouds.  That GMCs followed
these scalings became well established in a series of surveys and
catalogs of the CO emitting gas in the inner Milky Way
\citep[][S87]{sss85,dect86,syscw}.  Although there are several
algebraically equivalent forms for expressing these relations, the
most physically relevant relationships are generally agreed to be:
\begin{equation}
\sigma_v = \sigma_0 \left(\frac{R_e}{\mbox{1 pc}}\right)^{0.5}
\label{rdv}
\end{equation}
and 
\begin{equation}
M = \frac{5 \sigma_v^2 R_e}{\alpha G}.
\label{virial}
\end{equation}
For the inner Milky Way, $\sigma_0=0.6~$km~s$^{-1}$ and $\alpha
\approx 1.5$ (S87).  The first relationship is the size-line width
relationship and results from the supersonic turbulent motions in the
GMCs, as first suggested by \citet{larson}.  The constant $\alpha$ in
equation \ref{virial} is the virial parameter \citep{mckee-vt} and
since $\alpha < 2$ in the inner Milky Way the clouds are frequently
interpreted as being self-gravitating.  In Figure \ref{larson}, we
plot these two relationships for the 19 clouds in M31 with sufficient
signal to noise that their properties can be well-recovered in
interferometric observations.  We compare the relationships to those
found in the catalog of S87.

In both the size-line width relationship and the value of the virial
parameter, there is reasonably good agreement between the GMCs in M31
and the inner Milky Way GMCs of S87.  Fitting the relationship for the
19 clouds in M31 that meet the sensitivity criterion (\S\ref{fluxrec})
using the method of \citet{bces} gives a size-line width relationship
of
\begin{equation}
\log \sigma_v = (-0.5\pm 0.3) + (0.7\pm 0.2) \log R_e.
\end{equation}
For reference, a fit to the catalog of S87 using the same fitting
method, albeit without uncertainties in the cloud properties, gives 
\begin{equation}
\log \sigma_v = -0.28 + 0.55 \log R_e.
\end{equation}
Thus, the size-line width relationship for the molecular clouds is
indistinguishable from the clouds found in the inner Milky Way.  The
similarity between size-line width relationship for clouds in M31 and
the inner Milky Way has been reported on several occasions by other
authors \citep{vbb87,wilson_m31, iram-m31-orig, iram-m31,sheth},
though no previous study has reported an independent fit for the
clouds in M31 with well characterized uncertainties in the cloud
properties.

The virial parameter for clouds in M31 is constant as a function of
luminous mass.  The uncertainty-weighted, mean value of the virial
parameter for clouds in M31 is $\langle \alpha_{\mathrm{M31}} \rangle
= 2.0 \pm 0.3$.  For the sample of S87, $\langle \alpha_{\mathrm{MW}}
\rangle = 1.45$.  The difference between the virial parameters is not
statistically significant.  Moreover, any flux loss will affect the
luminous mass estimates more than the virial estimate (RL06), so our
measurement of the virial parameter is at worst an upper limit.  It is
unlikely that the true virial parameters of M31 cloud population
differs significantly from the Milky Way, provided the CO-to-H$_2$
conversion factor is the same for both systems.  Alternatively, by
assuming that the molecular clouds cataloged in this study are in the
same dynamical state as are clouds in the inner Milky Way, we can
justify our choice of CO-to-H$_2$ conversion factor.  Again, the
similarity of the virial parameter has been noted by several authors
in previous studies
\citep{lada_m31,iram-m31-orig,iram-m31,m31-zermatt}.  In particular,
the latter two studies find no systematic variation in the virial
parameter for clouds over a large range of galactocentric radius in
M31.

\section{The Angular Momentum Defect of GMCs}
\label{spang}

\subsection{Velocity Gradients}
In addition to macroscopic properties like line width, radius and
mass, interferometric observations of extragalactic GMCs can be used
to measure the velocity gradients and angular momenta of the clouds.
\citet[][RPEB]{rpeb03} analyzed the velocity gradients of GMCs in M33
to critically evaluate cloud formation theories.  In this section, we
analyze the velocity gradients for GMCs in M31 and compare our results
to those of RPEB and the analysis of the velocity gradients in GMCs in
the inner Milky Way by \citet{koda-gradient}.

We determine the velocity gradient by a least-squares fit of a plane
to the velocity centroid surface $v_c(\alpha,\delta)$ -- the velocity
centroid measured as a function of position across the cloud.  The
velocity centroid and its uncertainty are determined from the first
moment of velocity weighted by brightness.  The coefficients of the
fit determine the magnitude ($|\nabla v|$) and position angle
($\phi_\nabla$) of the velocity gradient for the molecular cloud
\citep[see][for details]{gbmf93}.  The results of the fits are listed
in Table \ref{m31-cloudprops}.  For purposes of this analysis, we have
included all clouds, since the velocity gradient is less affected by
the interferometric effects which bias the macroscopic properties
(\S\ref{fluxrec}).  The velocity gradient is derived from line
centroids and the first moments (centroids) are much less sensitive to
the low surface brightness wings of the line than are the second
moments.  We do exclude all measurements of the velocity gradient
where the uncertainty is larger than 0.15 km~s$^{-1}$~pc$^{-1}$.  This
upper limit was chosen from the maximum uncertainty for clouds with
high signal-to-noise ($T_{max}/\sigma_{rms}>10$), including a total of
44 clouds in this portion of the analysis.  Uncertainties in the
derived parameters are determined from the errors in the least-squares
parameters given the uncertainty in the centroid.

The magnitude of the velocity gradients range between 0 and 0.2
km~s$^{-1}$~pc$^{-1}$, in good agreement with both the values and
distribution found for clouds in M33 (RPEB) and the Milky Way
\citep{koda-gradient}.  We compare the magnitude of the velocity
gradient to the velocity gradient that would be expected from the
orbital motion of the gas in the galactic potential, ignoring
streaming motions in the spiral arms.  The local galactic velocity is
estimated from the tangent plane to a surface defined by the line of
sight projection of the galactic rotation velocity.  We use the
rotation curve of \citet{m31-rotcurve} which is a compilation of
previous CO and \ion{H}{1} studies using updated orientation
information.  On average, the rotation gradient of GMCs exceeds the
circular galactic gradient by a factor of 3.0.

We measured the difference between the position angle of the gradient
and that of the local galactic gradient expected from circular
rotation for each of the clouds.  While suggesting a slight alignment
with the galaxy, the distribution of this difference is not
significantly different from a random distribution according to a
two-sided KS test \citep[$P_{KS}=0.24$]{numrec}.  This is similar to
the work of \citet{koda-gradient} who found a random distribution of
position angles in the Milky Way GMCs but different from RPEB who
found significant alignment of the gradients with the galactic
rotation.

This analysis has assumed the decomposed data represent discrete
clouds with independent velocity fields, and the factor of three
difference between the magnitudes of the cloud and local galactic
gradients supports that model.  However, it is also possible that we
are inappropriately decomposing a continuous gas flow into clouds and
the large velocity gradients result from the change in the gas flowing
into the density wave.  In the presence of the radial motions that
such a flow implies, the local galactic velocity gradient will change
in magnitude by a factor of, at most,
$\sqrt{V_{r}^2+V_{\theta}^2}/V_{\theta}$ and in position angle by
$\tan^{-1} (V_{r}/V_{\theta})$ where $V_{r}$ and $V_{\theta}$ are the
radial and azimuthal components of the gas flow.  For gas in M31,
$V_{r}/V_{\theta}\sim 0.1$ \citep{braun-m31} so these changes will not
be significant and the factor of three excess of cloud gradient
magnitude over the local rotational gradient implies discontinuous
velocity fields.  Moreover, any changes should be systematic in
nature, and the large range of position angles actually observed
implies we are measuring the velocity gradients of discrete objects.


\subsection{Angular Momentum}

While the magnitudes of the velocity gradients of the clouds are
larger than the local rotation in the galaxy, the implied angular
momentum of the GMCs is significantly smaller than that of an
equivalent mass of atomic gas orbiting in the galactic gravitational
potential at the GMC's position.  The specific angular momentum for a
cloud of gas with velocity gradient magnitude $|\nabla v|$ is
$j_{GMC}=\beta |\nabla v| R_e^2$ where $\beta=0.4\pm 0.1$ for a wide
range of density distributions undergoing uniform rotation
\citep{p99}.  If the velocity gradient arises solely from the
turbulent motions in the cloud, $\beta$ is likely 2--3 times smaller
\citep{bb00}.

It is unlikely that the high galactic inclination affects the results.
The near-random distribution of the position angles implies that the
cloud angular momentum vectors are similarly close to random which
would require scaling the angular momentum up by a factor of $4/\pi$
on average (RPEB).  A turbulent origin for the velocity gradients
would not necessitate any correction.  Since any correction would be
small, we make no correction for the orientation of the clouds.

For the clouds in the velocity gradient analysis, the mean angular
momentum is $(36 \pm 3)$ pc km s$^{-1}$ for a solid body rotator and
2--3 times smaller for the angular momentum of turbulent flows,
comparable to the results for other systems
\citep[RPEB,][]{koda-gradient}.  We compare this to the angular
momentum of progenitor material for the GMCs. The angular momentum of
the gas in the galactic potential is given by RPEB \citep[equivalent
to][]{mestel}
\begin{equation}
j_{gal} = \eta \Delta R_{gal}^2 \frac{1}{R_{gal}} 
\frac{d}{dR_{gal}} (R_{gal} V_{\theta})
\end{equation}
where $\eta$ is a form factor equal to 1/4 for a contraction of a
uniform circular disk of material of radius $\Delta R_{gal}$.  The
length, $\Delta R_{gal}$, is the range of galactocentric radii over
which material is accumulated to form a GMC.  The minimum value of
$\Delta R_{gal}$ is set by the mass of the GMC under consideration and
the surface density of gas that could comprise the progenitor
material: $M_{GMC} = \pi \Sigma_{prog} \Delta R_{gal}^2$.  We take
$\Sigma_{prog}= 10 ~ M_{\odot}\mbox{ pc}^{-2}$ \citep{iram-m31-aa},
including their estimates of both atomic and molecular material since
small molecular clouds seen in CO may be progenitor material for the
larger GMCs.  From the measured masses of GMCs and locations of GMCs,
we calculate $\Delta R_{gal}$ and the galactic shear from the rotation
curve of \citet{m31-rotcurve}.  The mean value of $\langle
j_{gal}\rangle = (63\pm 5) \mbox{ pc km s}^{-1}$ for the clouds in the
sample.  On a cloud by cloud basis, the angular momentum of the
galactic material is only a factor of 1.7 larger than the that of the
GMCs, significantly smaller than the difference found in M33 where
$\langle j_{gal}/j_{GMC} \rangle = 5$.  The smaller margin is
primarily due to the lower shear in M31, reducing $j_{gal}$ by a
factor of 4 compared to that found for the clouds in M33 whereas the
velocity gradients are comparable in the two systems.  Similarly, the
cloud properties and gradients for clouds in the Milky Way are
comparable to the clouds in M31 \citep{koda-gradient}, but the shear
is much larger in the inner Milky Way compared to the 11-kpc arm of
M31.  Consequently, the discrepancy between the angular momentum of
progenitor material and the resulting GMCs is also larger in the inner
Milky Way than it is in M31 \citep[see][]{psp3}.  The angular momentum
defect for the M31 clouds is likely significant: the factor of 1.7
difference should be regarded as a minimum value since attributing the
velocity gradients to turbulence or accumulating the progenitor
material from a larger range of galactocentric radius will appreciably
increase the margin.

These results suggest a broad similarity between GMCs in M31 and the
other disk galaxies in the Local Group.  The primary goal of studying
the velocity structure within the GMCs is to relate their angular
momentum to that of the ISM as a whole to evaluate formation
mechanisms.  The angular momentum defect between GMCs and the ISM is
common among the studied systems \citep[][ RPEB]{psp3} and suggests
the importance of braking in molecular cloud formation.  \citet{kos03}
study the role of magnetic effects at removing angular momentum from
forming clouds and conclude that magnetic effects can reduce the
angular momentum to observed values.  Their unmagnetized simulations
show no significant reduction in cloud angular momentum, although the
density enhancements that represent proto-GMCs show
$j_{gal}/j_{gmc}\sim 2$, similar to the observed {\it minimum} value
found in M31.  The relatively low-shear environment does not
necessarily require magnetic braking to produce the observed cloud
angular momenta, though our measurement of $j_{gal}/j_{gmc}$ likely
significantly underestimates the true value.

\section{Galactic Environment and GMC Properties}
\label{environment}
With respect to their macroscopic properties and velocity gradients,
GMCs in the 11-kpc, star-forming arm in M31 appears indistinguishable
from the molecular clouds in the inner Milky Way. As our knowledge of
GMCs in other galaxies grows, it becomes more apparent that such
detailed agreement in the properties of GMCs is not universal and may
actually represent exceptional cases \citep{psp5}.  Given the
differences measured among extragalactic cloud populations, the most
pressing question is: what physical mechanisms regulate the observed
differences in GMCs properties?  These mechanisms must couple the GMCs
to the ISM on much larger scales, and future studies will examine the
nature of this coupling in more detail.

The similarity between GMCs in M31 and the Milky Way offers a useful
point of reference for studying the relationship between the galactic
ISM and GMCs.  Finding environmental properties that vary
significantly between the two systems while GMC properties remain
the same would imply those properties are irrelevant to the regulation
of molecular cloud properties.  Several features of the galactic
environment are thought to play a role in regulating the structure of
the cold ISM and the properties of GMCs.  Here, we review several of
these features and note the degree of similarity to the local Milky
Way:
\begin{itemize}

\item {\it Interstellar Radiation Field} --- 
The interstellar radiation field in M31 has been previously noted to
be poor in the ultraviolet relative to the Milky Way
\citep{cesarsky-m31}, though the M31 field may be significantly
enhanced in the star-forming spiral arm imaged in the CO observations.
Indeed, work by \citet{m31-pdr} study [\ion{C}{2}] emission in the
star forming arms at $R_{gal}\sim 10$~kpc and note that the far
ultraviolet field is consistent with being 100 times the local
interstellar value (i.e. $G_{0}=10^2$) around molecular clouds.  Thus,
the observed FUV deficiency likely does not characterize the molecule
rich region of the galaxy and the radiation environment is similar to
clouds in the Milky Way.

\item {\it Metallicity} ---  The gas phase metallicity from \ion{H}{2}
regions in M31 at $R_{gal}=10$ kpc is approximately
[O/H]-[O/H]$_{\odot}$=0.0 dex with significant scatter, up to $\sim$
0.2 dex \citep{m31-abund1,m31-abund2}.  The region shows no signs of
being significantly enriched or depleted relative to the solar
neighborhood.

\item {\it Dynamics} --- The Toomre $Q$ parameter is frequently used to
evaluate the stability of galactic disks with respect to the formation
of self-gravitating structures.  The parameter is defined as
\begin{equation}
Q\equiv \frac{\kappa \sigma_g}{\pi G \Sigma_g}
\end{equation}
where $\kappa$ is the epicyclic frequency, $\sigma_g$ is the
one-dimensional gas velocity dispersion and $\Sigma_g$ the surface
density of the gas in the disk.  Most star-forming regions of galactic
disks show $Q\approx 1.5$ \citep{mk01} and this region of M31 is no
exception.  For the rotation curve of \citet{m31-rotcurve},
$\kappa=0.03~\mbox{km s}^{-1}\mbox{ pc}^{-1}$.  We take
$\Sigma_g\approx 10~M_{\odot}$ \citep{iram-m31-aa}.  Finally, we take
$\sigma_g\approx 8\mbox{ km s}^{-1}$ from \citet{m31-unwin-sigmav}.
Combining these data give $Q\approx 1.8$, well within the range of
values observed in most galaxies and approximately the same value
found in the solar neighborhood.

\item {\it Midplane Density} --- We can also compare the midplane
particle density of this region in M31 to the solar neighborhood.  The
mean particle density is important for regulating thermal structure of
the ISM and the formation of the cold medium \citep{ism-wolfire}.
Adopting a gas scale height of 340 pc for this region of M31
\citep{braun-m31} gives $\langle n \rangle = 0.5\mbox{ cm}^{-3}$ in
the spiral arm region, but falling to $\langle n \rangle = 0.1 \mbox{
cm}^{-3}$ outside the arm.  These densities are smaller than that
found in the solar neighborhood \citep[$\langle n \rangle = 1.0 \mbox{
cm}^{-3}$,][]{ism-wolfire}.
\end{itemize}

The properties of the ISM in this region of M31 appear, at least
superficially, similar to those found in the solar neighborhood.
Hence, the derived similarity of molecular clouds in the two galaxies
is not surprising.  The only property that appears like it may be
significantly different between the two environments is the midplane
particle density in the neutral medium.  If the difference is
significant, the constancy of the GMC properties suggests that the
particle density may be important in regulating the fraction of
material found in the cold, neutral ISM but perhaps not the properties
of individual GMCs.  While this environmental similarity does not
isolate any irrelevant physical processes, noting the similarity in
both clouds and environments illustrates that similar galactic
environments do indeed produce similar GMCs.

\section{Summary and Conclusions}

This paper has presented BIMA millimeter interferometer observations
of $^{12}$CO($J=1\to 0$) emission from giant molecular clouds (GMCs)
along a spiral arm at $R_{gal}\approx 10$ kpc in M31.  The
observations were conducted in two parts: a low-resolution survey
using the compact (D) configuration of the array and a
high-resolution, follow-up study using a more extended (C)
configuration of the array.  All data from the study were segmented
and analyzed with the analysis algorithms of \citet{props}, which
correct the observational biases imprinted on the data by having
relatively low signal-to-noise and marginally resolving the GMCs.  The
C-array follow-up spans 7.4~kpc$^2$ finding 67 clouds.  We report the
following conclusions:

1. The GMC population is very similar to that found in the inner Milky
Way.  Both the size-line width relationship and distribution of virial
parameters are statistically indistinguishable from those derived from
the population of clouds in the inner Milky Way.  This confirms the
work of several authors \citep{vbb87,wilson_m31, iram-m31-orig,
iram-m31,sheth} while using techniques that correct for the biases
imprinted by using interferometric observations.

2. We measured the velocity gradients and angular momenta across the
GMCs.  Like the Milky Way and M33, there is significantly more angular
momentum in possible progenitor material than in the resulting GMCs,
though the discrepancy is smallest in M31 owing to the relatively low
galactic shear at large galactocentric radii.  All three systems
suggest braking mechanisms act during the process of cloud formation
to dissipate angular momentum.

3. The galactic environment in M31 where the GMCs are found is similar in
most respects that of the solar neighborhood and inner Milky Way.
That GMCs in both environments are similar is consistent with the idea
that the galactic environment regulates GMC properties.  In systems
where the galactic environment differs significantly from those
studied here, the GMC properties are found to be different
\citep[e.g.,][]{psp5}.  

\acknowledgements

This work was supported by a NASA GSRP fellowship and an NSF
postdoctoral fellowship (AST-0502605).  I am extremely grateful to
Rainer Beck, Leo Blitz, Michel Gu\'elin, Adam Leroy, Chris McKee, Eve
Ostriker and an anonymous referee for carefully reading the manuscript
and offering in-depth commentary.  Kartik Sheth kindly discussed his
current and future work on M31, allowing me to read a preprint of a
forthcoming paper.  This work made extensive use of the NASA
Extragalactic Database, the NASA Abstract Data Service, and the MAST
archive.  The molecular line observations used the BIMA millimeter
array which was supported, in part, by NSF grant AST-0228963 to the
Radio Astronomy Laboratory at U.C. Berkeley.


\end{document}